# Microwave band gap and cavity mode in spoof-insulator-spoof waveguide with multiscale structured surface

**Qiang Zhang[1], Jun Jun Xiao[1,*], Dezhuan Han[2], Fei Fei Qin[1], Xiao Ming Zhang[1] and Yong Yao[1]**

[1] College of Electronic and Information Engineering, Shenzhen Graduate School, Harbin Institute of Technology, Shenzhen 518055, People's Republic of China

[2] Department of Applied Physics, Chongqing University, Chongqing 400044, People's Republic of China

[*]E-mail: eiexiao@hitsz.edu.cn

**Abstract**

We propose a multiscale spoof-insulator-spoof (SIS) waveguide by introducing periodic geometry modulation in the wavelength scale to a SIS waveguide made of perfect electric conductor. The MSIS consists of multiple SIS subcells. The dispersion relationship of the fundamental guided mode of the spoof surface plasmon polaritons (SSPPs) is studied analytically within the small gap approximation. It is shown that the multiscale SIS possesses microwave band gap (MBG) due to the Bragg scattering. The "gap maps" in the design parameter space are provided. We demonstrate that the geometry of the subcells can efficiently adjust the effective refraction index of the elementary SIS and therefore further control the width and the position of the MBG. The results are in good agreement with numerical calculations by the finite element method (FEM). For finite-sized MSIS of given geometry in the millimeter scale, FEM calculations show that the first-order symmetric SSPP mode has zero transmission in the MBG within frequency range from 4.29 GHz to 5.1 GHz. A cavity mode is observed inside the gap at 4.58 GHz, which comes from a designer "point defect" in the multiscale SIS waveguide. Furthermore, ultrathin MSIS waveguides are shown to have both symmetric and antisymmetric modes with their own MBGs, respectively. The deep-subwavelength confinement and the great degree to control the propagation of SSPPs in such structures promise potential applications in miniaturized microwave device.

# 1. Introduction

The propagation characteristics of electromagnetic wave are related to not only the refraction index of the constituent media but also the spatial arrangement. Periodic modulation is one of the most commonly used schemes to control the behavior of electromagnetic waves in inhomogeneous media. Generally, a periodic structure may be considered as an effective medium or a Bragg structure, depending on the size of the lattice, e.g., the unit cell $P$. If $P$ is much smaller than the wavelength of the electromagnetic wave, i.e. $P \ll \lambda$, one can regard the whole structure as a homogenous medium with effective permittivity $\varepsilon_{eff}$ and permeability $\mu_{eff}$. Metamaterial is such an example whose $\varepsilon_{eff}$ and $\mu_{eff}$ can be engineered by carefully designing its periodic or random elementary scattering units (or meta-atoms) [1-3]. So far, various ($\varepsilon_{eff}$, $\mu_{eff}$) pair and wave dispersions that are not exist in nature materials have been achieved by metamaterials, for example, in negative refractive materials [4-6], zero refractive index materials [7-9], hyperbolic materials [10-12], and so on.

Recently, a new kind of plasmonic metamaterial made of semi-infinite perfect electric conductor (PEC) perforated with one-dimensional array of grooves or two-dimensional array of holes has been proposed. These microstructured PEC surfaces support the so called spoof surface plasmon polaritons (SSPPs) which have quite similar properties to the usual surface plasmons in noble metal nanostructures [13-17]. For example, localized spoof surface plasmon resonance (LSSPR) has been observed in microstructured PEC disk [18, 19]. The LSSPR can generate coherent phenomenon that are originally found in real plasmonic structures, such as Fano resonance in the heterodimer resonators [20-22]. The SSPP and LSSPR structures have additional geometrical degrees of freedom in designing the plasmon frequency which can be pushed down away from the visible, and enable THz and microwave surface plasmon modes that can find applications in THz sensing and microwave guiding [15, 16, 23, 24]. The grooves or the holes in the SSPP structures are much smaller than the wavelength. Therefore, they can be considered as an effective medium whose dielectric function has been proved to be a plasmonic form.

On the other hand, when $P$ is comparable to the wavelength, the Bragg scattering plays an important role and microwave band gap (MBG) or photonic band gap (PBG) emerges in the structure, as in the broadly studied photonic crystals [25-28]. In addition to the PBG of photonic crystals, the cavity mode associated to defects is also highly concerned in Bragg-type structures. As a matter of fact, many applications of photonic crystal are relevant to the PBG and the defect states, for example, as optical filter, omnidirectional reflector, and the high-Q photonic cavity [29-31].

In this context, the metamaterial concept and Bragg-type structure concept are commonly believed to be incompatible to each other. However, under certain circumstances, these two concepts are brought together, for example, in multiscale hyperbolic metamaterials [10] and in zero-n gap photonic crystal superlattice [32]. Both of them are realized by adding a wavelength-scale periodic superstructure to the metamaterials which already have substructures in the sub-wavelength scale. Such composite structure is also referred to as multiscale periodic structure. The multiscale structure has the virtues of both the metamaterial and the photonic crystal, thus providing more degrees of freedom to adjust the electromagnetic responses of the entire structure. It is noticed that the wavelength concerned in multiscale metamaterials is the effective wavelength determined by the geometry instead of by the constituent material. For example, multiscale multilayer hyperbolic metamaterials and hypercrystals are reported recently with band gap existing in both the wave vector and the frequency domains [10, 33]. In these multiscale structures, the size $P$ of the superstructure is at the wavelength scale of the high-$k$ volume SPPs ( $P \sim \lambda_{\mathrm{SPP}}$ ) that is much smaller than the corresponding vacuum wavelength.

In this paper, we design a multiscale structure by adding a wavelength-scale periodic modulation to a one dimensional (1D) spoof-insulator-spoof (SIS) waveguide [34-37]. We focus on the MBG in the multiscale SIS (MSIS) and show how the substructure can become a flexible dimension to adjust the MBG, with respect to both the central position and the width. The small gap approximation (SGA) is introduced to analyze the dispersion of the symmetric mode in 1D SIS and further use it to examine "gap maps" of the 1D MSIS. We numerically study the transmission spectra in finite-sized MSIS, focusing on the MBG and the cavity mode. In addition, the band structure of an ultrathin MSIS is also investigated by numerical calculation.

## 2. Dispersion and gap maps

Before discussing the proposed MSIS, we firstly review the properties of electromagnetic wave propagation in a SIS waveguide. Generally, the unit cell of a SIS consists of two counter-face PEC surfaces with 1D arrays of grooves drilled in them, as schematically shown in figure 1(*a*). The structure is assumed of thickness $t$ in the $z$-direction. In fact, each side of the waveguide walls of the SIS structure is a PEC grating surface that sustains SSPPs [13]. The dispersion relation of SSPPs of two dimensional (2D) (i.e., $t \to \infty$) single PEC grating surface has been studied by mode expansion method and transfer matrix approach in the effective medium approximation [14, 38, 39]. In the mode expansion, all the high-order diffraction orders by the PEC grating can be neglected if the widths of the SIS unit cell $d \ll \lambda_0$ ($\lambda_0$ the wavelength in air). By matching the boundary conditions and monitoring the divergence of the reflection coefficient, one obtains the close form dispersion relation of a single PEC grating surface:

$$\beta^2 = \left(\frac{a}{d}\right)^2 k_0^2 \tan^2(k_0 h) + k_0^2 . \qquad (1)$$

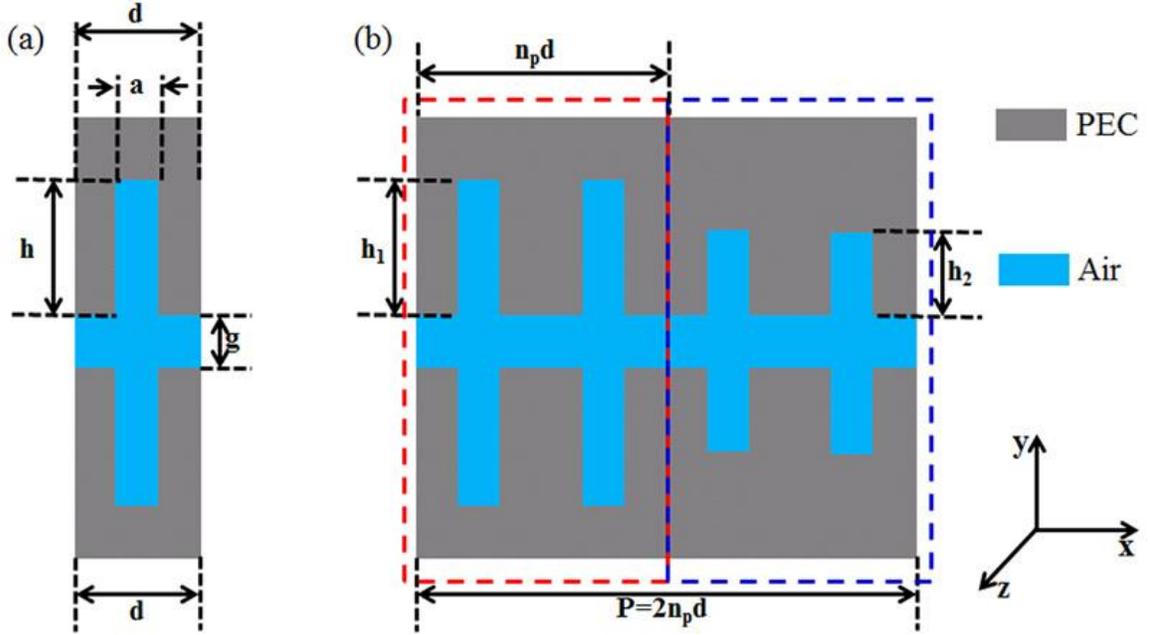

**Figure 1.** (Color online) Geometry of the 1D MSIS waveguide. (*a*) The elementary cell which consists of two PEC surfaces with 1D array of grooves drilled in them. The groove width *a* and the length *d* are much smaller than the operating wavelength. (*b*) Unit cell of the MSIS containing two subcells, marked by the dashed boxes correspondingly. The number of the elementary cells in each SIS subcell is $n_p$ ($n_p = 2$ here in this figure) and the depth of the grooves are $h_1$ and $h_2$, respectively. The rest geometry parameters in the MSIS are the same as those in (*a*).

Here *a* and *h* are respectively the width and the depth of the groove, $\beta$ is the propagating constant, and $k_0 = 2\pi/\lambda_0$ is the free space wave number [14]. Equation (1) holds for transverse magnetic polarization (i.e., $E_z = 0$). In the effective medium description, the single PEC grating surface is regarded as an anisotropic homogenous medium with $\varepsilon_x = a/d$ and $\varepsilon_y = \varepsilon_z = \infty$. We note that the same expression of equation (1) can be obtained by the transfer matrix method based on the effective medium theory. As to the SIS, Kats et al obtained the rigorous closed form dispersion in a model film of six layers [36]. Same as in a metal-insulator-metal waveguide, the SSPPs in the SIS waveguide can be categorized into symmetric and antisymmetric modes [40]. Their corresponding dispersion relations are respectively given by

$$\frac{\sqrt{\beta^2-k_0^2}}{k_0}\tanh(\frac{g}{2}\sqrt{\beta^2-k_0^2}) = \frac{a}{d}\tan(k_0 h), \tag{2}$$

$$\frac{\sqrt{\beta^2-k_0^2}}{k_0}\coth(\frac{g}{2}\sqrt{\beta^2-k_0^2}) = \frac{a}{d}\tan(k_0 h). \tag{3}$$

The $n$-th order groove resonance frequency is $\omega_n = n\pi c/2h$ and this groove resonance frequency is also the asymptotic frequency of the $n$-th SSPPs mode in the SIS. Although there are multiple modes of different order in the SIS, we only concern the fundamental one ($n=1$) in this work for simplicity. In SGA $g\sqrt{\beta^2 - k_0^2}/2 \ll 1$, we can take the first-order Taylor expansion to the hyperbolic tangent function in equation (2). The SGA actually requires small gap width $g$ and/or small deviation between the vector and the light line. In other words, $\beta^2 - k_0^2$ in equation (2) should be relatively small. We note that similar approximation was often utilized to study real SPPs in metal-insulator-metal waveguides [41-43]. In this SGA, equation (2) reads

$$\beta^2 = k_0^2 + \frac{2a}{dg}k_0 \tan(k_0 h). \tag{4}$$

Figure 2 shows the dispersion diagram of the fundamental symmetry mode for different $g$ given by equations (2) and (4). Since equation (4) is obtained from equation (2) by the Taylor approximation, the errors of this approximation may yield discrepancy between the dispersion curves obtained by equation (2) and equation (4). The geometry parameters are $d = 5$ mm, $h = 10$ mm and $a = 2$ mm. It is seen that results by equation (4) can indeed be good approximations to those by equation (2). For example, for $g \leq 1$ mm the dispersion obtained by equation (2) agrees well with that obtained by equation (4) at propagating constant up to $\beta = \pi/d$ (~ 0.628 mm$^{-1}$). For very large gap, e.g., $g = 50$ mm, the apparent differences between equations (2) and (4) can be seen for large $\beta$ (see blue lines and circles in figure 2). However, in the frequency range that the dispersion is closed to the light line (green dashed line), the SGA is still acceptable. From equation (4), the effective refraction index $n_{eff}$ of the symmetric SSPPs mode is

$$n_{eff} \equiv \frac{\beta}{k_0} = \sqrt{1 + \frac{2a}{dgk_0}\tan(k_0 h)}. \tag{5}$$

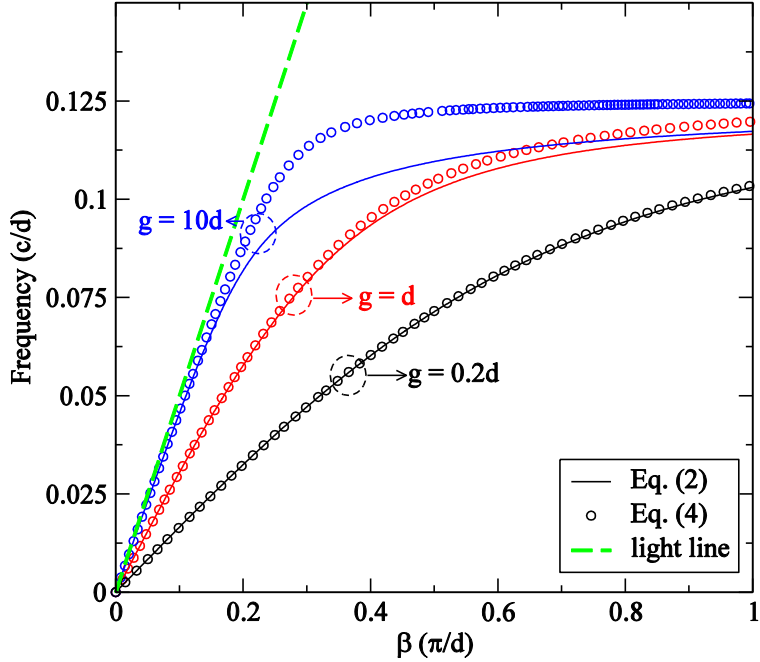

**Figure 2.** (Color online) Dispersion diagram of the first-order symmetric SSPPs mode in the SIS for different gap $g$. Results are obtained by equation (2) (line) and equation (4) (symbol), respectively.

It is obvious that $n_{eff}$ of the symmetric SSPPs modes propagating in the SIS can be engineered by the slot width $g$, the duty circle $a/d$, as well as the depth of the groove $h$.

Next, we consider adding a periodic modulation to the SIS in the operating wavelength scale. It is stressed that the wavelength here refers to the wavelength of the SSPPs $\lambda_{sspp}$ in the SIS, instead of that in air $\lambda_0$. The wavelength-scale periodic modulation makes the whole structure a MSIS waveguide, as shown in figure 1(b). The supercell of the MSIS contains two SIS units with different geometry and the individual SIS [see figure 1(a)] now becomes a subcell of the MSIS waveguide. It is straightforward by equation (5) that any geometry difference (e.g., $g$, $a/d$, and $h$) between the subcells of these two SIS structures can result in different $n_{eff}$. Here in this paper we consider the case of different $h_1$ and $h_2$ in the two SIS. The groove depth $h$ controls the spoof plasmon frequency $\omega_a = \pi c/2h$ and therefore can dramatically affect the whole dispersion curve of the SIS. Nevertheless, the following discussions can be applied to the cases of the remaining geometry parameters. The dispersion equation of the symmetric mode in the MSIS is obtained from the Bloch theorem [10, 44]

$$\cos(qP) = \cos(\beta_1 d_1)\cos(\beta_2 d_2) - \frac{1}{2}(\frac{\beta_1}{\beta_2} + \frac{\beta_2}{\beta_1})\sin(\beta_1 d_1)\sin(\beta_2 d_2), \tag{6}$$

where $q$ is the Bloch wave vector. In the proposed MSIS structure, we set $d_1 = d_2 = n_p d$, and therefore $P = 2n_p d$. In equation (6), $\beta_i (i = 1, 2)$ are the effective propagating constant in the individual subcells, dictated rigorously by equation (2) or approximately by equation (4) in the SGA. Figure 3 shows the gap maps of the MSIS for different geometry parameters of the subcell. Each panel of figure 3 represents the case of varying one of the geometry parameters with default values $d = 5\,\text{mm}, n_p = 2$, $g = 0.25P$, $a/d = 0.4$, $h_1 = 0.5P$, and $h_2/h_1 = 0.5$. Here in each SIS $n_p \geq 2$ to make sure $P = 2n_p d = 20\,\text{mm}$ is comparable to the effective SSPP wavelength in the subcells. If we set $n_p = 1$, the MBG disappears and the MSIS in this case becomes a dual band structure [17]. These default values are chosen to ensure that the waveguide operates in the microwave regime and the SGA is valid. Certainly, we can set other default values to tune the bands and gaps to the terahertz regime. The gap maps provide detailed information about the position and the width of the MBG. The filled regions in figure 3 indicate the MBGs wherein the microwave cannot go through the MSIS. Here we mainly focus on the lowest gray region which denotes the MBG of the first-order symmetric SSPP mode. The red thick lines are the numerically calculated first and second band edges at the irreducible Brillouin zone boundary ($q = \pi/P$), respectively. They are obtained by eigenfrequency analysis with a finite element method (FEM) solver (COMSOL Multiphysics). Notice that in the numerical calculations, the metal structure is modeled as ideal PEC. The theoretical predictions agree quite well with the numerical ones, enabling convenient engineering of the forbidden gaps in the MSIS waveguide.

Figure 3(*a*) shows how the width of the slot $g$ affects the MBG of the MSIS. It is seen that when $g = 0$, the slot is completely closed and the band gap disappears (actually all waves cannot propagate in this case). As g increases from zero slowly, the MBG shifts to the higher frequency with its width broadened. The shifting speed slows down for increased $g$, indicating that the MBG is more sensitive to the $g$ variation in the smaller gap regime. In figure 3(*a*), the maximum $g$ is set to be $P$ since the SGA fails for very large $g$. Though, the deviations between the numerical results (red curves) and the theoretical ones (filled region) are still noticeable for $g > 0.5P$. The duty circle $a/d$ of subcell is another

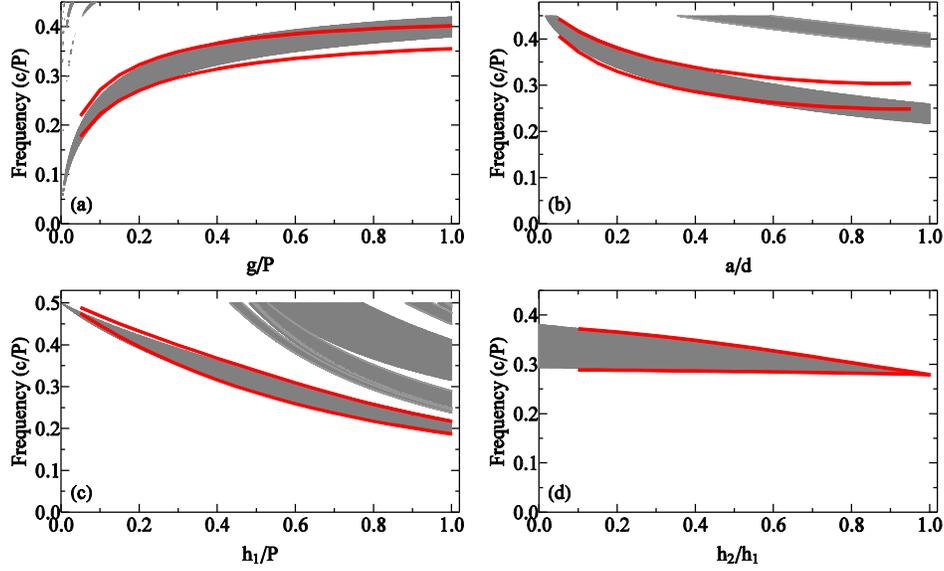

**Figure 3.** (Color online) "Gap maps" for the geometry parameters of the subcell with varying (*a*) slot size $g$, (*b*) duty circle of the grooves $a/d$, (*c*) depth $h_1$ of the grooves of the left SIS, and (*d*) depth contrast $h_2/h_1$ of the grooves in the two SIS subcells. The gray filled regions represent the MBGs obtained by the SGA and the red thick lines are the boundaries of the MBG by FEM calculations.

important parameter that affects the gap map, as shown in figure 3(*b*). There are two extreme cases of $a/d = 0$ and $a/d = 1$. The former one represents the situation that the structure transforms from MSIS to planar PEC slot waveguide whose fundamental mode dispersion is exactly the light line [45]. Therefore $a/d = 0$ defines the upper limit of the edge of the MBG which is $f^+_{lim} = cq/2\pi = 0.5c/P$, predictable by the light line. As long as $a/d \neq 0$, a forbidden gap emerges, with its position descends from $f^+_{lim}$ to that in the other extreme case. That is, when $a/d \rightarrow 1$, the width of the PEC wall of the grooves approaches zero and the dispersion curve deviates the farthest away from the light line, according to equation (4). Therefore, as the duty circle $a/d$ gradually increases, the mismatch between the numerical calculation (red lines) and the SGA (gray regions) becomes heavier, as shown in figure 3(b).

We then turn to the gap map for the depth ($h_1$, $h_2$) of the grooves. We firstly consider changing $h_1$ while keeping their ratio constant as $h_2 = 0.5h_1$. The corresponding gap map is plotted in figure 3(*c*). The structure becomes a PEC planar waveguide at $h_1 = 0$ and the upper limit of the band gap is at $f^+_{lim} = 0.5c/P$. As $h_1$ increases, the band gap shows up in a lower frequency range. Interestingly, it is also seen in figure 3(*c*) that the width of the band gap has the maximum for $h_1 \sim 0.5P$. The reason remains elusive. We note that for large groove depth $h_1$, higher-order bands emerge and the higher-order band gaps are also observed in figure 3(*c*).

Finally, we examine how the contrast of the depths $h_2/h_1$ in the two SIS subcells can control the band gap. The results are shown in figure 3(*d*). It is seen that as $h_2/h_1$ varies from 0 to 1, the upper boundary of the MBG approaches to the lower one until the MBG completely vanishes at $h_2=h_1$. Notice that the lower boundary of the MBG remains relatively inert to the change of $h_2/h_1$. This shows that large groove depth difference leads to broader MBG, reasonably as the effective refractive index contrast grows. Further notice that when $h_2=0$, the MSIS become a combination of the SIS and the planar PEC waveguides. This structure shows a very broad MBG.

## 3. Microwave transmission in finite-sized samples

In this section, we focus on the transmission of SSPP waves propagating in finite-sized MSIS samples. Figure 4(*a*) shows the band structure of a MSIS with concrete geometry parameters. The MBG falls in the GHz band, as predicted by figure 3. Only the fundamental mode dispersion is involved here. Again, the dispersion is obtained by equation (6) in the SGA (black solid line) and by FEM eigenfrequency calculations (black circle). It is seen that the results by both methods agree well except for the slight difference at the high-frequency portion of the second band. To compare with the individual SIS, the dispersions of the SIS waveguides of $h_1=10$ mm (blue line with downward triangle) and $h_2=5$ mm (red line with upright-triangles) are superimposed in figure 4(*a*). We see that the MBG falls in the frequency range from 4.29 GHz to 5.08 GHz, right between the band edges at the zone boundary of the two corresponding SIS waveguides. It is stressed that those modes with wave vector resides in the light cone (green dash line) are not SSPPs but the common guided waves. Figure 4(*b*) shows the transmission spectrum of a MSIS with 15 units (total length $L_{tot}=300$ mm) obtained via FEM simulation with a port excitation. The frequency range of zero transmission is perfectly consistent with the MBG shown in figure 4(*a*). The ripple in the transmission spectrum of figure 4(*b*) comes from the multiple Bragg reflections. To eliminate the reflection between the excitation port and the SIS waveguide, impendence matching devices that take advantage of the gradient grooves have been proposed [46]. In contrast to the single scale SIS waveguide which can be used as a broadband band pass filter [47], the MSIS demonstrated here can service as a band rejection filter within the frequency of the MBG. For example, excitation at $f = 4.69$ GHz is prohibited to go through the structure, as clearly seen in the inserts of figure 4(*b*). Figures 4(c) and 4(d) show the band structure and the corresponding transmission spectrum for different $h_1/h_2$. It is seen that the band gap position can be easily adjusted by $h_1/h_2$ and the larger contrast between $h_1$ and $h_2$ leads to a broader MBG. Figure 5 shows the electric field along the central line inside one supercell for band edge eigenmodes, i.e., at $f = 4.29$ GHz and $f = 5.08$ GHz. The insets of figure 5 show the corresponding field patterns. It is shown that the electric field concentrates on the right (left) SIS subcell for $f = 4.29$ GHz (5.08 GHz). This is quite similar to a 1D photonic crystal wherein the electromagnetic energy accumulates in the high (low) refraction index layer for modes at the top (bottom) of the PBG [27].

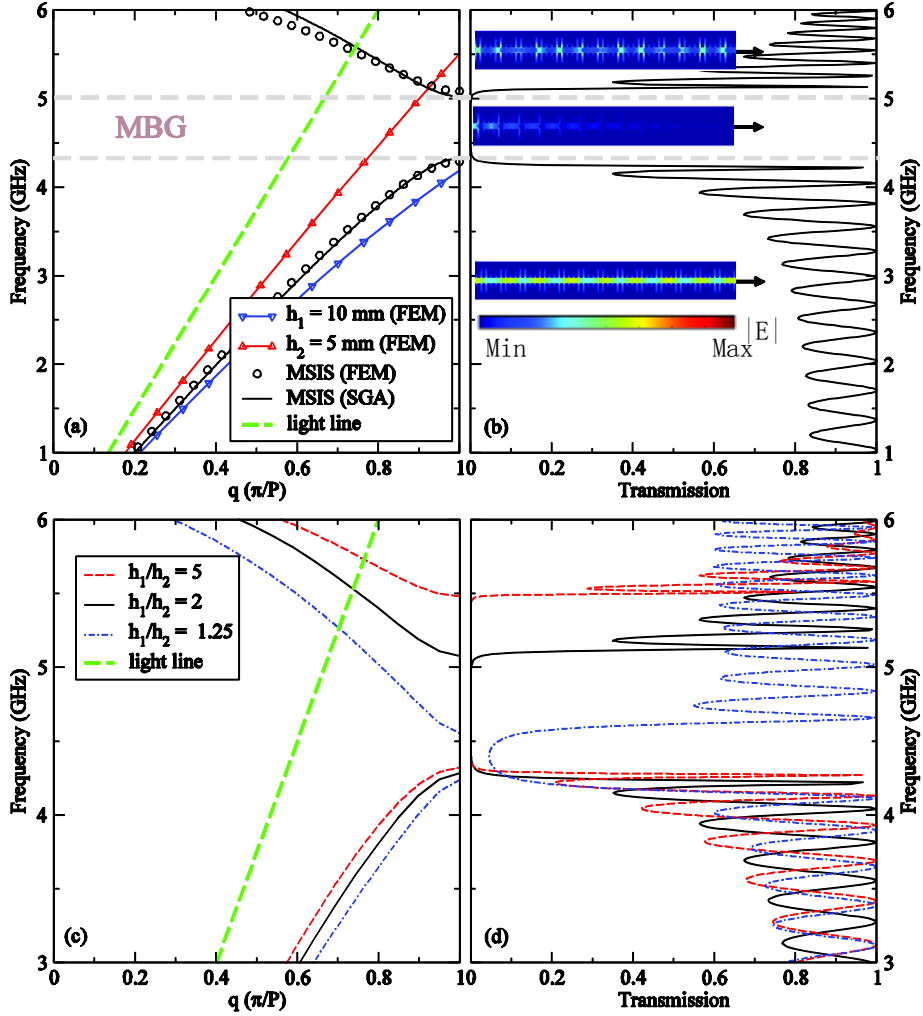

**Figure 4.** (*a*) (Color online) Dispersion diagram of a specific MSIS and the individual SIS with $h_1 = 10$ mm and $h_2 = 5$ mm. The MBG of the MSIS lies in the frequency range from 4.29 GHz to 5.08 GHz, as marked by the horizontal gray dashed lines. (*b*) Transmission spectrum for a finite-sized MSIS of 15 periods via a port excitation. Inserts in (*b*) are the electric fields at frequencies of $f = 3$ GHz, 4.69 GHz, and 5.55 GHz, respectively marked by the black arrows. (c) Dispersion diagrams for different $h_1/h_2$. The MBGs of the MSIS obviously shift as $h_1/h_2$ changed. (d) Transmission spectrum for the corresponding MSIS in (c).

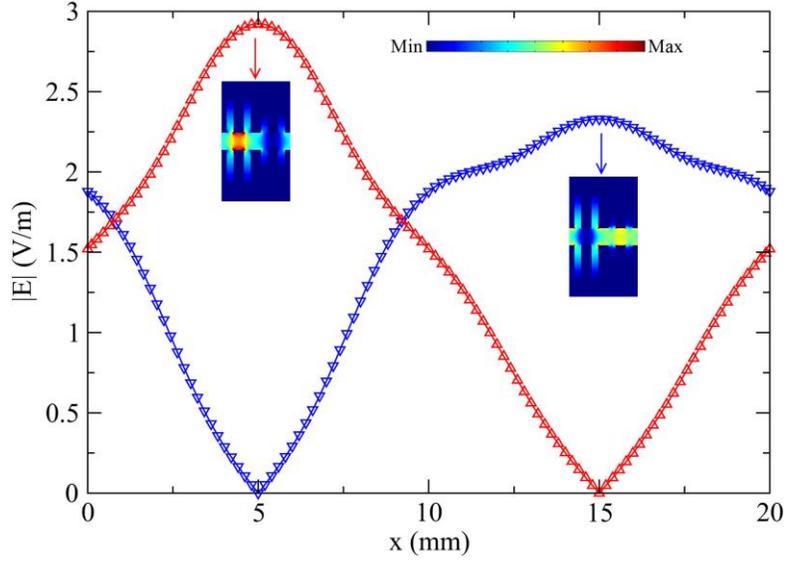

**Figure 5.** (Color online) Electric field amplitude in one unit cell of the MSIS waveguide for the MBG boundary frequencies of $f = 4.29$ GHz (line with '△') and $f = 5.08$ GHz (line with '▽'). The inserts show their corresponding 2D field distribution.

In photonic crystals, defect mode is often used to generate high Q cavity for a lot of applications such as lasing, strongly coupled cavities, and channel add-drop filter [48-50]. In view that the MSIS can be regarded as a microwave analogue to photonic crystal, we expect cavity modes if a defect is engineered in the structure. For that purpose, we deliberately eliminate the groove portion in the middle period [see the schematic representation in the insert of figure 6(a)] in the finite-sized seen that a very narrow transmission peak at frequency $f = 4.578$ GHz appears in the bang gap region. For the first order SSPP mode considered here, the frequency of the cavity mode shall be smaller than the asymptotic frequency, i.e. the spoof plasmon frequency $f_a = c/4h_1 = 7.5$ GHz. In principle, the frequency of the cavity mode is independent of the defect position which has been also verified in our calculation (figures not shown). Under the port excitation at the left terminal, the electric field distribution along the waveguide at this frequency is plotted in figures 6(b) and 6(c). It is seen that the electric field reaches its maximum near the defect position ($x \approx 8P$) and decay exponentially [see the envelope in figure 6(b)] towards the two sides of the waveguide. The quality factor of the cavity mode is estimated to be $Q \approx 460$. We note that dual cavities or multiple cavities are also readily incorporated into the MSIS waveguides and the order and disordered cavity effects can be explored based on such system.

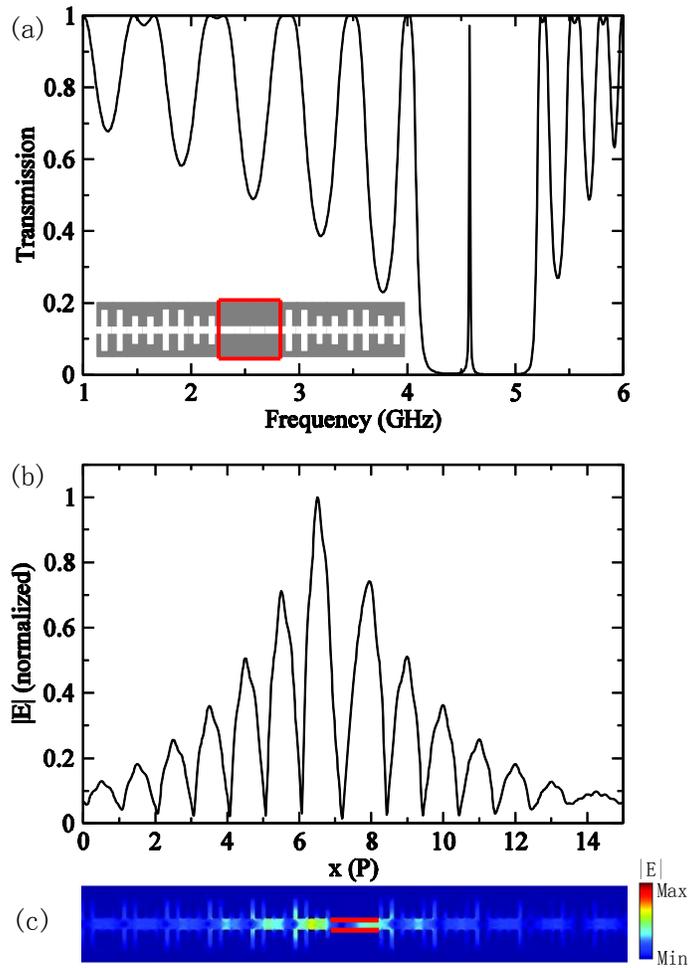

**Figure 6.** (Color online) (a) Transmission spectrum of MSIS with a defect. The inset shows the schematic representation of the eliminated portion of the structure. (b) Electric field $|E|$ of the cavity mode excitation at $f = 4.578$ GHz along the waveguide central line. (c) 2D field pattern $|E|$ for the cavity mode excitation at $f = 4.578$ GHz. The red lines mark the position of the defect.

## 4. Band structure of ultrathin MSIS structure

In previous sections, the waveguides are assumed to be of 2D. However, in practical applications, the proposed structure must be fabricated in a metal film of finite thickness. In general, the thickness of such structure is quite small, i.e., an ultrathin version of such waveguides could be employed. There are a lot of studies on single-side PEC grating surfaces (namely, SSPP slabs) [15, 17, 23]. However, no experimental studies have been reported on SIS or MSIS structures. Figure 7(a) shows the dispersion relations of a single side SSPP slab. The numerical dispersion curve of the finite thickness structure compares favorably with Xiang et al [51], and is spectrally below that of the corresponding infinite sample. Figure 7(*b*) shows the calculated dispersions of the MSIS studied in figure 4, but with finite thicknesses $t = 0.1$ mm, $t = 1$ mm and $t = 5$ mm. Here we only present the dispersion of the first-order SSPPs and modes out of the light cone lie in the frequency range from 0 to 7 GHz. From figure 7(b) we see that there are four first-order SSPP bands for all the three thicknesses. Two of them correspond to the symmetric modes (red lines) while the other two correspond to the antisymmetric modes (blue lines). More importantly, the antisymmetic bands are more sensitive to the thickness variation.

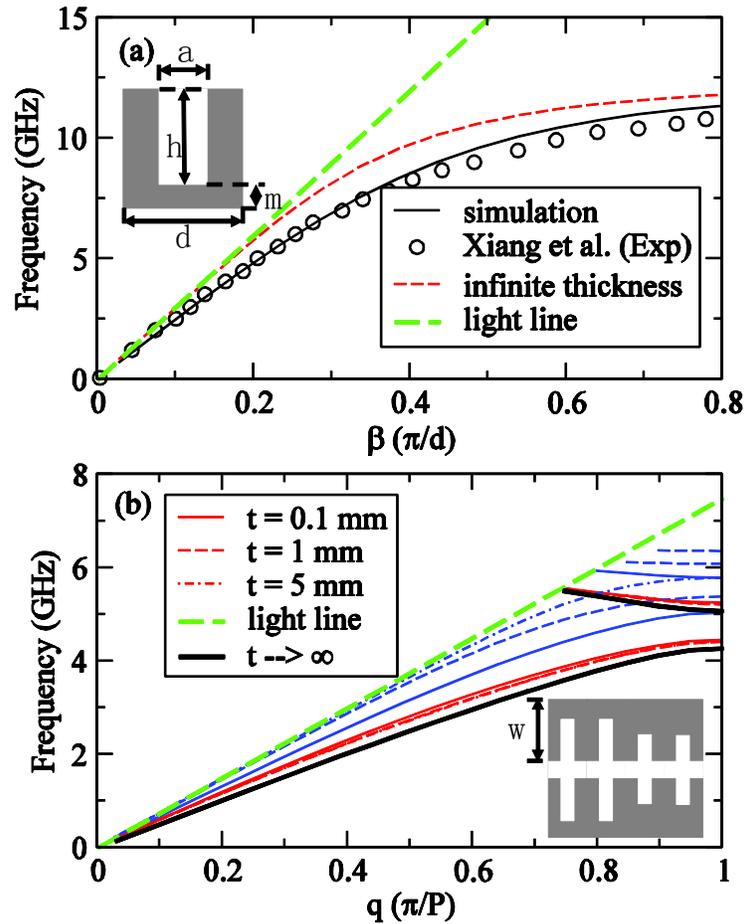

**Figure 7.** (color online) (a) Dispersion relations of a single side SSPPs slab obtained by both simulation and experimental measurement. The insert shows the unite cell with $a = 2$ mm, $d = 5$ mm, $h = 5$ mm, $m = 3$ mm and the thickness $t = 0.05$ mm. (b) Band structures of the MSIS waveguides with different thicknesses. The red lines represent symmetric modes and the blue lines are for antisymmetric modes. The geometry parameters are the same as in figure 4 except for $w = 15$ mm.

As an example, we label the four bands of the sample with thickness $t = 0.1$ mm as 'S1', 'AS1', 'S2', and 'AS2' in figure 8(a), consequently from the low frequency to the high one. In order to clarify the corresponding features of the four bands, we have plotted in figure 8(*b*) to figure 8(e) the field pattern $H_z$ in the $xy$ plane ($z = 0$), for modes at the zone boundary as labeled by '**b**', '**c**', '**d**', and '**e**' in figure 8(*a*). We have utilized the field mirror symmetry (symmetry or antisymmetry) with respect to the $x$-axis and the parity (even or odd) symmetry with respect to the y-axis in the individual subcells, each containing two pairs of grooves. Specifically, the magnetic fields along the $y$-axis of symmetric mode are in-phase [see figures 8(*b*) and 8(*d*)] and those of antisymmetric ones are out-of-phase [see figures 8(*c*) and 8(*e*)]. On the other hand, the symmetry characteristics in the $x$ direction of the magnetic field inside each subcell can be regarded as even or odd. For example, for the S1 band, the magnetic fields in the left SIS subcell are "even" and those in the right SIS subcell are "odd". We therefore label it as '**EO**'. While for the S2 band, the fields are "odd" in the left SIS subcell but "even" in the right SIS, labeled as '**OE**'. Therefore, the dispersion lines S1 and S2 in figure 8(*a*) correspond to symmetric SSPP modes but of different combination of parity symmetry. While the AS1 and AS2 curves are associated with antisymmetric modes. The dispersion diagram shown in figure 8(*a*) indicates that both the symmetric mode and the antisymmetric mode can generate their own MBGs. Interestingly, the symmetric and antisymmetric modes do not necessarily have completely overlapped MBGs. Therefore if we consider them simultaneously, a true complete MBG only exist in the intersection region between their respective MBGs. In this case, however, since the defect in the finite thickness structure is no longer a closed cavity and may radiate to the third dimension, the cavity mode would be subtle and demand a systematic and careful study. In practice, there must be a substrate to support the waveguide. In most cases the substrate structure (e.g., foams) is usually transparency with respect to the working microwave and has slight influences on the spoof SPPs. As we mentioned before, a transition device can be employed to reduce the insertion loss between the feeding part and the SSPP waveguide. Such transition devices have been successfully designed and applied to the SSPPs slabs [46, 52, 53]. Since the MSIS waveguide is fundamentally similar to the SSPPs slab, it is reasonable to believe that it is highly possible to achieve efficient transmission for the MSIS, using a carefully designed transition device, even with presence of a substrate.

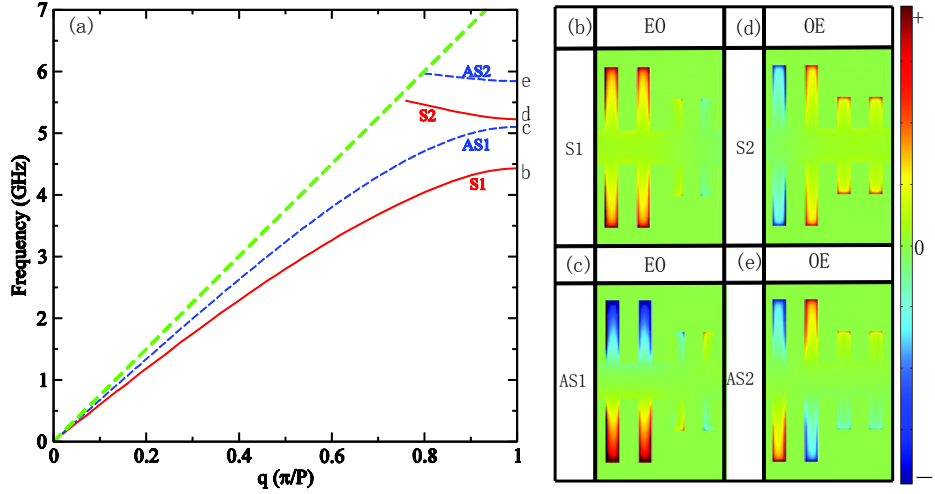

**Figure 8.** (Color online) (*a*) Dispersion diagram of three-dimensional MSIS with thickness of 0.1 mm. The lines S1 (AS1) and S2 (AS2) correspond to the symmetric (antisymmetric) SSPP modes. (*b*)-(*e*) $H_z$ distributions at $q = \pi/P$ for each branch at the corresponding points labeled in (*a*) with '**b**', '**c**', '**d**', and '**e**'.

## 5. Conclusion

In summary, we have demonstrated that multiscale subwavelength spoof-insulator-spoof waveguide possesses resonant and propagation features of both metamaterial and photonic crystal. As far as PEC assumption holds, the results can be scaled from microwave to THz, as well as to the near-infrared. The geometry of the subcells in such waveguide provides flexible control over the dispersion of SSPP waves travelling inside. Particularly, the width and the position of the forbidden gap can be extended all the way down to GHz range. We demonstrate that the transmission properties and the field distribution of microwave confined in the MSIS are very similar to those of light wave in 1D photonic crystal. In addition, we show that the cavity mode of a structural defect can be easily engineered. Even when electronic plasmon comes into play, the structured surfaces can still provide the extra freedom. The MSIS waveguide can therefore be used to control the propagation characteristics of electromagnetic wave in deep-subwavelength scale and may find applications in highly integrated microwave and plasmonic devices.

**Acknowledgements**

This work was supported by NSFC (11274083, 11304038, and 11374223), and SZMSTP (Nos. JCYJ20120613114137248, KQCX20120801093710373, and 2011PTZZ048). J.J.X is supported by Natural Scientific Research Innovation Foundation in Harbin Institute of Technology (No. HIT.NSRIF.2010131). We acknowledge assistance from the National Supercomputing Center in Shenzhen.